# Switching Transition in a Resource Exchange Model on Graphs

Shreeman Auromahima[1], Sitangshu Bikas Santra[2], Biplab Bose[1,*]

[1]Department of Biosciences and Bioengineering, Indian Institute of Technology Guwahati, Assam, India

[2]Department of Physics, Indian Institute of Technology Guwahati, Assam, India

*email: biplabbose@iitg.ac.in

In this work, we investigate a simple nonequilibrium system with many interconnected, open subsystems, each exchanging a globally conserved resource with an external reserve. The system is represented by a random graph, where nodes represent the subsystems connected through edges. At each time step, a randomly selected node gains a token (*i.e*, a resource) from the reserve with probability (1–$p$) or loses a token to the reserve with probability $p$. When a node loses a token, its neighbors also lose a token each. This asymmetric token exchange breaks the detailed balance. We investigate the steady state behavior of our model for different types of random graphs: graphs without edges, regular graphs, Erdős–Rényi, and Barabási–Albert graphs. In all cases, the system exhibits a sharp, switch-like transition between a token-saturated state and an empty state. When the control parameter $p$ is below a critical threshold, almost all tokens accumulate on the graph. Furthermore, in a non-regular graph, most tokens accumulate or condense on nodes of minimum degree. A slight increase in $p$ beyond the threshold drains almost all the tokens from the graph. This switching transition results from the interplay between drift and the conservation of tokens. However, the position of the critical threshold and the behavior at the transition zone depend on graph topology.

**Keywords:** state transition; nonequilibrium; graph; stochastic process

## I. INTRODUCTION

A hallmark of a complex system is that it is composed of a large number of interacting subsystems. Often, these subsystems are open, exchanging materials and energy with the environment. Examples of such systems are ubiquitous. Consider the example of a tissue in a multicellular organism or a consortium of bacteria. Individual cells in the tissue are open systems and exchange material and energy with their surroundings. They also interact with each other, often through specific molecular processes and the competition/sharing of resources. The same applies to bacteria in a consortium. Similarly, individual firms in the economy are open subsystems that exchange goods, capital, and labor with the market and also interact with one another directly or indirectly.

Additionally, there could be some drive or regulation that may affect the dynamics of the system and turn it into a nonequilibrium one. The interplay of these features, openness, interactions, and drive/regulation, may give rise to a plethora of emergent behaviors in a complex system. One such emergent behavior is the state transition, where the system moves from one particular state to another with a change in a control parameter [1]. Such state transitions can be smooth, but sometimes they are drastic or even catastrophic [2,3]. Distinct and sharp transitions between states are well known across natural and social systems. In ecology, gradual changes in environmental factors, such as nutrient availability and temperature, are known to drive discrete transitions between two states, resulting in the loss of specific organisms within an ecosystem [4]. Similar transitions are noted in the financial market [5], opinion dynamics, and other social dynamics [6].

State transition in nonequilibrium systems has been investigated using various toy models [7,8]. Several lattice-based models with local interactions or rules and an external drive, such as the driven lattice gas model, have been investigated [7,9]. These systems retain the basic equilibrium framework of an energy function that governs local updates. However, some form of external drive is used to bias the dynamics and push the system out of detailed balance. Some nonequilibrium models, such as directed percolation [10] and ASEP/TASEP [11,12], do not involve any notion of energy, Hamiltonian, or temperature. Instead, in these systems, the dynamics are fully specified by stochastic update rules, such as asymmetric hopping rates [10], directional biases [11], or probability thresholds [13].

In certain nonequilibrium models, the system is open and exchanges particles with an external reserve. One such model is the open boundary ASEP [14], where particles are injected at one end of the system and removed from the other with fixed probabilities. This system exhibits a nonequilibrium steady state and displays a phase transition between high- and low-density states.

Chemical reaction processes where molecules are continuously supplied and removed are open systems and can reach nonequilibrium steady states. Many such systems undergo bifurcation, where the system transitions from one distinct steady state to another when a control parameter exceeds a critical value. Bifurcation is akin to a phase transition. One such system is the Schlögl model [15], where simple coupled reaction systems exhibit first-order and second-order transitions.

In the current work, we developed a minimal nonequilibrium model of state transition in a system composed of open and interconnected subsystems. We used a graph to represent the interaction between subsystems, where nodes represent the subsystems and edges represent the connections or interactions between them. Each node stochastically exchanges a token (representing matter or resource) with a reserve (representing the environment). The total number of tokens, on the graph plus in the reserve, is constant. At every discrete time step, a randomly selected node either loses or gains a token with specific probabilities. The probability of token loss acts as a system-wide control parameter that regulates the drift of tokens.

Nodes do not exchange tokens among themselves, but affect each other's token exchange with the reserve. When a node loses a token, its neighbors also lose one token each. Therefore, the token exchange is asymmetric − at a given time step, the graph can gain only one token from the reserve but can lose more than one token to it. In other words, this is a model of individual gain vs collective loss.

We investigated the steady state behavior of this model. We demonstrate that this simple model undergoes a state transition in which the graph shifts from a token-saturated state to an empty state through a sharp, switch-like transition. We investigated both analytically and numerically the origin of this state transition and identified the connection between the position of the critical threshold and the graph's topology. Interestingly, we also observed a condensation transition-type phenomenon in the same system, where all tokens accumulate in only a handful of nodes.

## II. THE MODEL

We present a stochastic token exchange model on random graphs. The model consists of two components: a graph $G$ with $N$ nodes and $E$ edges, and a reserve R. The graph represents a system, and its nodes are subsystems. The nodes of the graph and the reserve hold tokens. Tokens are identical. These tokens are exchanged between the nodes of the graph and the reserve following the rules described below. The total number of tokens in the model is conserved. At any time, the tokens present on all nodes, together with those in the reserve, sum to a constant $T$, and generally, $T \gg N$.

At each discrete time step, one node is selected uniformly at random from the graph. With probability $p$, the selected node and each of its first neighbors lose one token, provided they each have at least one token. Otherwise, the selected node receives one token from the reserve, provided the reserve has at least one token. No other nodes gain tokens at this step. All tokens lost by the nodes at this step are transferred to the reserve.

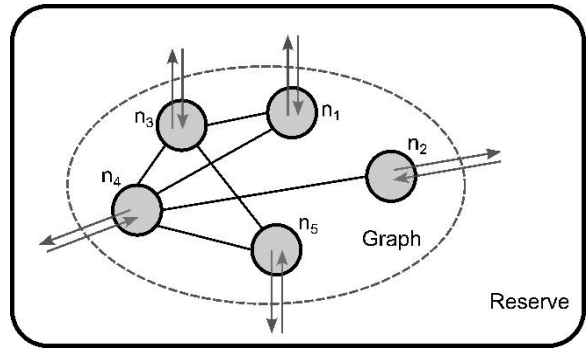

Figure 1. Schematic representation of the model. The model consists of two components – a graph (the ellipse with the dotted line) and a reserve (the box surrounding the ellipse). The graph consists of nodes (filled circles) and edges (lines) connecting the nodes. In the figure, the graph has $N = 5$ nodes and $E = 6$ edges. Nodes are labelled from $n_1$ to $n_5$. Each node can exchange tokens with the reserve as represented by arrows. The rules of token exchange are described in the model section. Tokens are not shown. The total number of tokens on the nodes and the reserve is conserved.

As an illustration, consider the model shown in Fig. 1. Suppose that at a particular time step, node $n_1$ is selected at random. It has two neighbors, $n_3$ and $n_4$. Now, two mutually exclusive events can happen. With probability $p$, nodes $n_1$, $n_3$, and $n_4$ will lose one token each, provided each of them has at least one token. If the selected node $n_1$ has no tokens, it does not lose any; nevertheless, $n_3$ and $n_4$ still lose one token each, provided they have tokens to lose. These tokens will go to the reserve. Otherwise, with probability $(1-p)$, $n_1$ will receive a token from the reserve if the reserve has at least one token. Note that the neighbors of $n_1$ do not gain any token in this step.

We investigated the steady state behavior of this model. Since the dynamics is stochastic, the total number of tokens on the graph, $T_G$, and in the reserve, $T_R$, fluctuate over time. We say a steady state is reached when the average values of $T_G$ and $T_R$ become independent of time. The fraction of tokens on the graph at the steady state, $f_G = T_G/T$, is the order parameter for the model, and $p$ is the control parameter.

We investigated our model for various types of graphs. To begin with, we analyze the behavior of the model when the graph has *N* nodes but no edges (*i.e.,* node degree $k = 0$). So, nodes have no neighbors. This system is simple enough to investigate analytically and serves as a reference for our subsequent studies.

Subsequently, we investigated the behavior for three types of random connected graphs− regular, Barabási-Albert (BA), and Erdős–Rényi (ER) graphs. In a regular graph, all nodes have the same degree. That means all of them have the same number of neighbors. So, the system is homogeneous.

ER graphs are canonical in the field of network analysis and widely used to compare different graph-based models [16]. The degree distribution of an ER graph follows the Poisson distribution. Therefore, most of the nodes have a degree close to the mean degree. These graphs have a very low clustering coefficient and do not have any hubs. On the contrary, in BA graphs, most nodes have the lowest degree, as the degree distribution follows a power law distribution [16]. Additionally, these graphs have hubs and exhibit degree-dependent clustering. Many real-life networks, such as metabolic networks, protein-protein interaction networks, social networks, and the World Wide Web, have similar properties [16].

For these three random graphs, we investigated our model through numerical simulation. In numerical simulations, if not mentioned otherwise, $N = 5000$, $T = 1000N$, and we started the simulations with all tokens in the reserve. The simulation for a given set of parameter values was repeated independently multiple times to calculate the ensemble average of steady state behavior.

## III. Results

### A. Behavior for a graph without any edges

Consider our model for a graph with *N* nodes but no edges. Since nodes are not connected, they are independent, and they have no neighbors. In the event of token loss, only the randomly selected node loses a token.

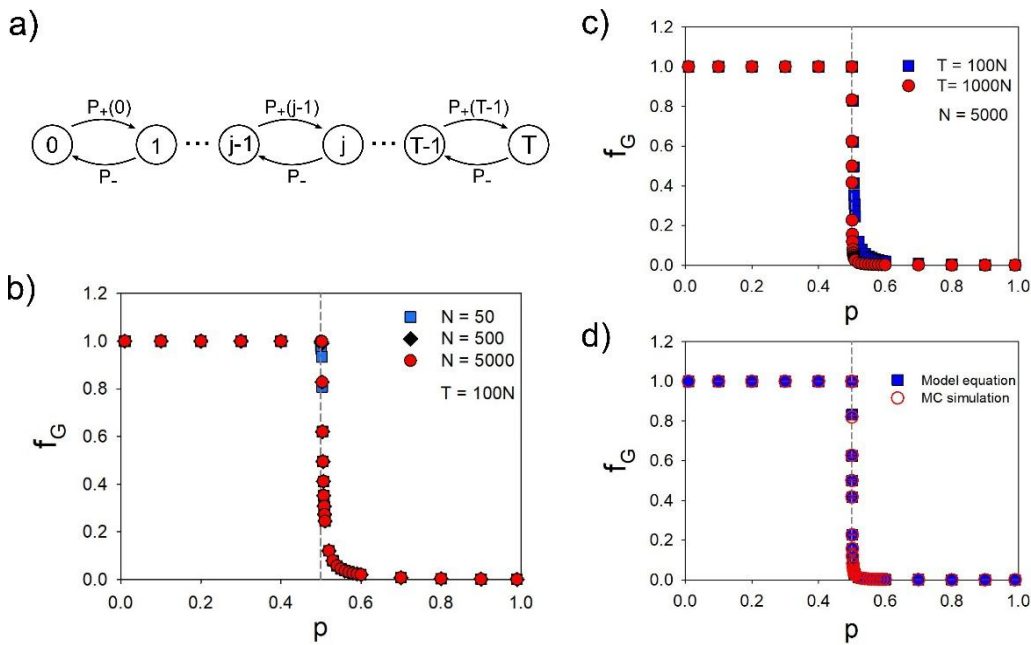

Figure 2. Steady state behavior for a graph without edges. a) State change diagram for the reserve. In the *j*-th state, the reserve holds *j* tokens. $j = 0,1,\ldots,T$. $P_+(j)$ and $P_-$ are the probabilities for token gain and loss, respectively. The fraction of tokens on the graph ($f_G$) at steady state in different cases is shown in (b), (c), and (d). b) $f_G$ vs *p* data for different numbers of nodes (*N*). c) Effect of the number of total tokens (*T*) on $f_G$ vs *p* behavior. Data in (b) and (c) were generated using the analytically derived Eq.(11) for $f_G$. d) Comparison of the analytical result with the result of the MC simulation. *N* = 5000 and *T* = 1000*N*. Model equation: data using Eq. (11). The vertical dotted lines in plots indicate the critical threshold $p_c = 0.5$.

We want to derive the fraction of tokens present on the graph ($f_G$) at the steady state. However, following the number of tokens in the reserve is analytically easier. As the total token is conserved, we can calculate $f_G$ from the number of tokens present in the reserve at the steady state. We use this approach in the following derivation.

At every time step, the reserve either loses a token, gains a token, or remains unchanged. The token dynamics of the reserve is equivalent to a finite discrete-time death-birth process with partially reflecting boundaries [17] as shown in Fig. 2a.

Let the reserve have $j$ tokens at a time step. $j = 0, 1, \ldots, T$. The reserve will receive a token from a randomly selected node with probability $p$, provided that the node has at least one token. Let $q_j$ be the probability that the selected node has at least one token. Then the probability of token gain by the reserve,

$$P_+(j) = q_j p \frac{1}{N} \quad (1)$$

If the reserve has $j$ tokens, the graph has $(T-j)$ tokens distributed over $N$ nodes. The number of ways this could be achieved is $\Omega_j = {}^{T-j+N-1}C_{N-1}$. Similarly, the number of ways $(T-j)$ tokens can be distributed over $(N-1)$ nodes, leaving the selected node empty, is $\Omega_j^0 = {}^{T-j+N-2}C_{N-2}$. As all these configurations are equally likely, the probability that the selected node has at least one token,

$$q_j = 1 - \frac{\Omega_j^0}{\Omega_j} = 1 - \frac{{}^{T-j+N-2}C_{N-2}}{{}^{T-j+N-1}C_{N-1}} = \frac{T-j}{T-j+N-1} \quad (2)$$

At a given time step, a randomly selected node can gain a token from the reserve with probability $(1-p)$. Therefore, the probability of a token loss from the reserve is,

$$P_- = (1-p)\frac{1}{N} \quad (3)$$

Let $\pi_j$ and $\pi_{j-1}$ denote the probabilities that the reserve has $j$ and $j-1$ tokens, respectively. Following the principle of detailed balance, at the steady state,

$$P_- \pi_j = P_+(j-1) \pi_{j-1}, \text{ for } j = 1, \cdots, T \quad (4)$$

Using Eq. (1) and (3) in Eq. (4)

$$\pi_j = q_{j-1}\left(\frac{p}{1-p}\right) \pi_{j-1} = q_{j-1} \beta \pi_{j-1}, \text{ for } j = 1, \cdots, T \quad (5)$$

Here, $\beta = p/(1-p)$. Using the Eq. (5) repeatedly, we get

$$\pi_j = q_{j-1} \beta \pi_{j-1} = \beta^j \pi_0 \prod_{i=0}^{j-1} q_i = Q_j \beta^j \pi_0 \quad (6)$$

Here, $\pi_0$ is the probability that the reserve has no tokens. Using, Eq. (2),

$$Q_j = \prod_{i=0}^{j-1} q_i = \prod_{i=0}^{j-1}\left(\frac{T-i}{T-i+N-1}\right) = \frac{{}^{T+N-j-1}C_{N-1}}{{}^{T+N-1}C_{N-1}} \quad (7)$$

As $\sum_{i=0}^{T} \pi_i = 1$, using Eq. (6) we get

$\sum_{i=0}^{T} \pi_i = \pi_0 \sum_{i=0}^{T} Q_i \beta^i = 1$. Hence,

$$\pi_0 = \frac{1}{\sum_{i=0}^{T} Q_i \beta^i} \quad (8)$$

Now we rewrite Eq. (6) using Eq. (7) and (8),

$$\pi_j = Q_j \beta^j \pi_0 = \frac{Q_j \beta^j}{\sum_{i=0}^{T} Q_i \beta^i} = \frac{\left({}^{T+N-j-1}C_{N-1}\right)\beta^j}{Z} \quad (9)$$

Here, $Z = \sum_{i=0}^{T} \left({}^{T+N-i-1}C_{N-1}\right)\beta^i$.

We use Eq. (9), to calculate the average number of tokens in the reserve at the steady state,

$$\langle j \rangle = \sum_{j=0}^{T} j \pi_j = \frac{1}{Z} \sum_{j=0}^{T} j \left({}^{T+N-j-1}C_{N-1}\right)\beta^j \quad (10)$$

Therefore, the fraction of tokens on the graph at the steady state,

$$f_G = 1 - \frac{\langle j \rangle}{T} = 1 - \frac{1}{TZ} \sum_{j=0}^{T} j \left({}^{T+N-j-1}C_{N-1}\right)\beta^j \quad (11)$$

Now, consider $p = 0.5$. Therefore, $\beta = 1$. In this case, Eq. (10) simplifies to

$$\langle j \rangle = \frac{1}{Z} \sum_{j=0}^{T} j \left({}^{T+N-j-1}C_{N-1}\right) \quad (12)$$

Through algebraic rearrangements, reindexing, and using the hockey-stick identity, it can be shown that

$$Z = \sum_{i=0}^{T} \left({}^{T+N-i-1}C_{N-1}\right)\beta^i = \sum_{i=0}^{T} \left({}^{T+N-i-1}C_{N-1}\right) = {}^{T+N}C_T \quad (13)$$

$$\sum_{j=0}^{T} j \left({}^{T+N-j-1}C_{N-1}\right) = T \left({}^{T+N}C_T\right) - N \left({}^{T+N}C_{T-1}\right) \quad (14)$$

Using, Eq. (13) and (14), we get a simpler form of Eq. (12),

$$\langle j \rangle = \frac{T \left({}^{T+N}C_T\right) - N \left({}^{T+N}C_{T-1}\right)}{{}^{T+N}C_T} = T\left(1 - \frac{N}{N+1}\right) \quad (15)$$

Therefore, when $p = 0.5$ the fraction of tokens on the graph at the steady state,

$$f_G = 1 - \frac{\langle j \rangle}{T} = \frac{N}{N+1} \quad (16)$$

So, when $p = 0.5$ and $N \gg 1$, then $f_G \approx 1$, and almost all the tokens will be on the graph.

However, what would happen when $p \neq 0.5$? Let us investigate the expected behavior of $f_G$ for $p \neq 0.5$ using the idea of average token drift for each node.

Notice that the graph has no edges. So, all nodes are independent and identical. Let $j$ tokens be in the reserve. When the reserve is not empty ($j > 0$), the average drift of tokens for a randomly selected node is,

$$\Delta = (1-p) - pq_j = 1 - (1+q_j)p \quad (17)$$

As mentioned earlier, $q_j$ is the probability that the selected node has at least one token, and $0 \leq q_j \leq 1$.

When, $p < 0.5$, Eq. (17) gives $\Delta > 0$. So, there will be a net drift of tokens from the reserve to the graph. Consequently, the reserve will be pushed towards $j = 0$ and spend considerable time there. So, at the steady state, most of the tokens will be on the graph, and $f_G \approx 1$.

Now consider $p > 0.5$. When $p > 0.5$ and $q_j > (1-p)/p$, then $\Delta < 0$. There will be a net drift of tokens from the graph to the reserve. That reduces $T_G$ and $q_j$. However, when $p > 0.5$ and $q_j < (1-p)/p$, then $\Delta > 0$ and there will be a net drift of tokens to the graph. That will increase both $T_G$ and $q_j$. These two opposing dynamics will drive the system to a balance or steady state where $\Delta = 0$ and $q_j = (1-p)/p$.

Using $q_j = (1-p)/p$ and $j = T - T_G$ in Eq. (2), we get

$$T_G = \frac{(N-1)(1-p)}{(2p-1)} \qquad (18)$$

Therefore, when $p > 0.5$, fraction of tokens on the graph at the steady state,

$$f_G = \frac{T_G}{T} = \frac{(N-1)(1-p)}{(2p-1)T} \approx \left(\frac{N}{T}\right)\left(\frac{1-p}{2p-1}\right), \quad \text{for } N \gg 1 \qquad (19)$$

Eq. (19) implies that when $p > 0.5$, $f_G$ decreases with an increase in $p$. Consider, $T \gg N \gg 1$. If $p$ is slightly larger than 0.5, $f_G$ stays close to one due to the boundary constraint $T_G \leq T$. However, as $p$ moves further away from 0.5, $f_G$ falls sharply and approaches values of order $N/T$ that are close to zero.

So, the drift-based analysis shows that when $p < 0.5$, $f_G \approx 1$, but when $p > 0.5$ $f_G$ drops toward zero. From Eq. (16), we know at $p = 0.5$, $f_G \approx 1$. Therefore, $p = 0.5$ is the critical threshold ($p_c$) for an edgeless graph.

Now, we confirm this behavior by numerically evaluating Eq. (11) for different values of $p$, $N$, and $T$. Note that Eq. (11) gives us the value of $f_G$ for any value of $p$, $0 \leq p \leq 1$, without any approximation. In Fig. 2b, we show the $f_G$ vs $p$ data for graphs with different numbers of nodes. As shown in the figure, the system has a sharp, switch-like transition at $p_c = 0.5$. As $p$ increases beyond this critical threshold, the graph jumps from the token saturated state ($f_G \approx 1$) to the empty state ($f_G \approx 0$) through a zone of very sharp transition. The stiffness of this transition increases with the total number of tokens (Fig. 2c).

We compared these results with the results obtained from our model's Monte Carlo simulation. As shown in Fig. 2d, the simulation result matches the result obtained using Eq. (11).

Therefore, in brief, when the nodes have no neighbors, our model exhibits a sharp, switch-like state transition, where the system jumps from a token-saturated state to an empty state as the control parameter $p$ increases beyond the critical threshold $p_c = 0.5$.

**B. Behavior for random graphs with edges**

When a graph has edges, the dynamics of token gain remains the same, but the dynamics of token loss changes. In the event of token loss, the selected node and all of its neighbors lose one token each, provided they have at least one token each. That should increase the rate of token loss and should affect the critical threshold that we identified in the last section.

Let us focus on the gain or loss of a token by a single node at one step. The probability that, in one step, a randomly selected node will gain a token,

$$P_g = (1-p) q_r \frac{1}{N} \qquad (20)$$

Here $q_r$ is the probability that the reserve has at least one token.

A node can lose a token in two cases: either it has been selected randomly, or any of its neighbors has been selected. Therefore, the probability that a node will lose a token in one step,

$$P_l = (1+k_i) p q_i \frac{1}{N} \qquad (21)$$

Here, $k_i$ is the degree of the node, and $q_i$ is the probability that the node has at least one token.

So, the average change in token numbers for that node is

$$\Delta_i = P_g - P_l = (1-p)\frac{q_r}{N} - (1+k_i)\frac{pq_i}{N} \qquad (22)$$

The node will not lose a token when
$$(1-p)q_r \geq (1+k_i) p q_i \qquad (23)$$

Let's consider that the selected node has at least one token, so that it has a chance of losing it. So, $q_i = 1$ and we know $q_r \leq 1$. Following the inequality (23), a necessary condition for a node with a degree $k_i$ to hold tokens is $k_i \leq \left(\frac{1}{p} - 2\right)$. In other words, when $p > 1/(k_i + 2)$, a node with a degree $k_i$ will not be able to hold tokens.

Let the minimum degree in a graph be $k_{min}$. When $p > 1/(k_{min} + 2)$, the loss of tokens will dominate over the gain for all nodes of the graph. There will be a net drift of tokens from the graph to the reserve. When $p < 1/(k_{min} + 2)$, nodes with degree $k_i \leq \left(\frac{1}{p} - 2\right)$ will be able to hold tokens, and there will be net drift of tokens from the reserve to the graph. Therefore, for a graph with minimum degree $k_{min}$, $p_c = 1/(k_{min} + 2)$ is the critical threshold.

In a graph without edges, all nodes have zero degree. So, $k_{min} = 0$ and accordingly, $p_c = 1/(k_{min}+2) = 1/2$. We derived the same $p_c$ in the last section.

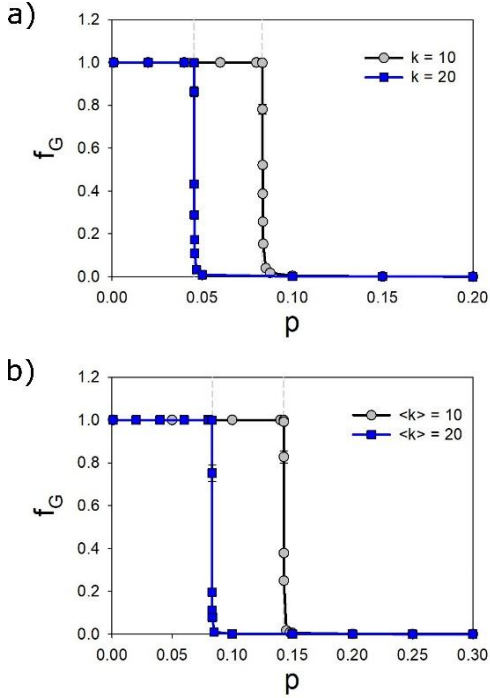

Figure 3: Fraction of tokens on regular and BA graphs at steady state. a) $f_G$ vs $p$ data for two regular graphs with degrees 10 and 20. b) $f_G$ vs $p$ data for two BA graphs with mean degree 10 and 20. The minimum degrees for these two BA graphs are 5 and 10, respectively. The dotted vertical lines on both graphs mark the critical threshold $p_c$. Averages of multiple independent simulations are shown, and the error bars are mostly smaller than the symbols.

In a random regular graph, all nodes have the same degree $k$. Therefore, the critical threshold for a regular graph would be $p_c = 1/(k+2)$. We simulated our model for two random regular graphs with degrees 10 and 20. The data is shown in Fig. 3a. For these two graphs, the critical thresholds are 1/12 and 1/22, respectively. As expected, in both graphs, for $p < p_c$ almost all the tokens are on the graph at steady state ($f_G \approx 1$). Beyond $p_c$ tokens drain off very fast, and $f_G$ drops to zero sharply with increasing $p$.

Fig. 3b shows the $f_G$ vs $p$ behavior for two random graphs generated with the BA algorithm [18]. The degree distribution of these two BA graphs follows the power law. The $k_{min}$ for these two graphs are 5 and 10. Accordingly, we expect that the critical thresholds should be $p_c = 1/(k_{min}+2) = 1/7$ and 1/12, respectively. As shown in Fig. 3b, in both cases for $p < p_c$, most tokens accumulate on the graph with $f_G \approx 1$. As $p$ crosses $p_c$, $f_G$ falls sharply to zero.

So, the overall behavior of these two types of random connected graphs is the same as for a graph without edges. Up to the critical threshold $p_c$, there is a net positive drift of tokens to the graph, accumulating almost all the tokens on the graph at the steady state. However, when $p > p_c$, there is a net negative drift and $f_G$ falls sharply to zero. The only difference is that for a connected graph, $k_{min}$ decides the $p_c$.

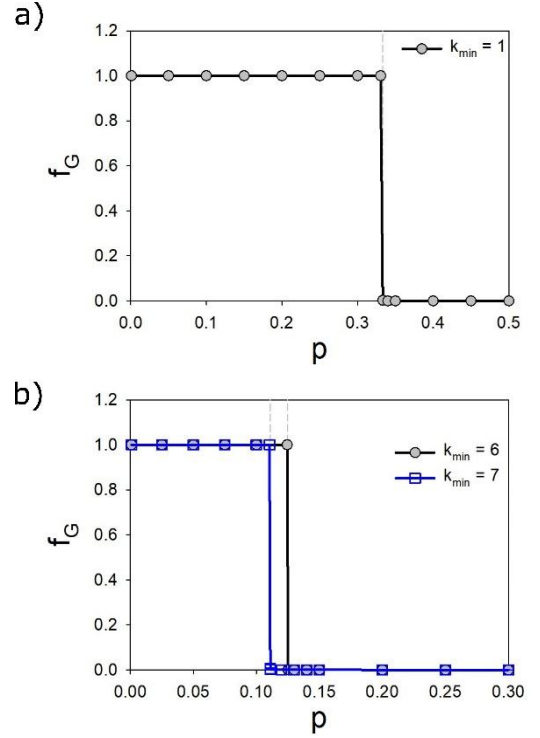

Figure 4: Fraction of tokens on ER graphs at steady state. Mean degree $\langle k \rangle$ is 10 and 20 in (a) and (b), respectively. In (a) minimum degree $k_{min} = 1$. However, in (b), two types of ER graphs with $\langle k \rangle = 20$ but two different $k_{min}$ are investigated. The dotted vertical lines on both the graphs mark the $p_c$. Averages of multiple independent simulations are shown, and the error bars are smaller than the symbols.

The effect of $k_{min}$ is more evident when we investigate the behavior of ER-type random graphs [19]. The degree distribution of these graphs follows a Poisson distribution [16], and the minimum-degree nodes appear with the least frequency. In numerical simulations, the $k_{min}$ of an ER graph varies from one realization to another, and such variation depends upon the number of nodes and the mean degree $\langle k \rangle$. In our simulations, we found that in all repeats the $k_{min}$ was one for ER graphs with $\langle k \rangle = 10$. Accordingly, the $p_c = 1/3$, and a sharp switch-like transition from $f_G \approx 1$ to $f_G \approx 0$ happens at that $p$ (Fig. 4a).

However, for $\langle k \rangle = 20$, $k_{\min}$ varied widely. For example, $k_{\min}$ was 5, 6, and 7 in ~21%, ~42% and ~27% cases, respectively. Therefore, we simulated ER graphs with $k_{\min} = 6$ and $k_{\min} = 7$, for $\langle k \rangle = 20$ separately. In both cases, we observed the sharp switch-like transition. However, as shown in Fig. 4b, $f_G$ changes sharply at two different threshold values of $p$ for these two graphs. These threshold values correspond to $p_c = 1/8$ and $p_c = 1/9$ for $k_{\min} = 6$ and $k_{\min} = 7$, respectively.

In essence, $f_G$ vs $p$ has a sharp switch-like behavior for all connected graphs, and the topological feature $k_{\min}$ decides the critical value of $p$ for this behavior. However, the behavior of ER graphs in the transition zone is different from the other. From the $f_G$ vs $p$ plots in Fig. 3 and 4, it is evident that, unlike regular and BA graphs, the transition from $f_G = 1$ to 0, in ER graphs is more like discrete jumps. We investigate this behavior in the following section.

**C. Token distribution and transition behavior**

First, we investigate the distribution of tokens in a regular graph with a degree $k \geq 0$ at steady state. In this case, all nodes are equivalent. Let $\{n_0, n_1, \cdots, n_j, \cdots, n_{T_G}\}$ be a configuration of the graph at the steady state, where $n_j$ is the number of nodes with $j$ tokens, such that $\sum_{j=0}^{T_G} n_j = N$ and $\sum_{j=0}^{T_G} j n_j = T_G$. At steady state, the system explores all microscopic configurations with equal probability. Therefore, the probability of observing a particular configuration is proportional to its multiplicity,

$$W = \frac{N!}{\prod_{j=0}^{T_G} n_j!}$$

The most probable distribution of tokens will maximize $W$ under the constraints of the total nodes and the total number of tokens on graph. We use the Lagrange multiplier method to maximise $\ln W$ (equivalently $W$) with the constraints $\sum_{j=0}^{T_G} n_j = N$ and $\sum_{j=0}^{T_G} j n_j = T_G$. The Lagrangian is,

$$\mathcal{L} = \ln W - \alpha \left( \sum_{j=0}^{T_G} n_j - N \right) - \mu \left( \sum_{j=0}^{T_G} j n_j - T_G \right) \quad (24)$$

Using the sterling approximation for $\ln W$ and setting $\frac{\partial \mathcal{L}}{\partial n_j} = 0$ we get,

$$n_j = e^{-(1+\alpha)} e^{-\mu j} \quad (25)$$

When, $T_G \gg 1$, using Eq. (25) and solving for the constraints $\sum_{j=0}^{T_G} n_j = N$ and $\sum_{j=0}^{T_G} j n_j = T_G$, it can be shown that $e^{-(1+\alpha)} = N(1-e^{-\mu})$ and $\mu = \ln(1 + N/T_G)$. Therefore, the probability of nodes with $j$ tokens,

$$P(j) = \frac{n_j}{N} = (1-e^{-\mu}) e^{-\mu j} \quad (26)$$

Let $\theta = (1-e^{-\mu}) = \frac{N}{T_G + N}$. We rewrite Eq. (26) as,

$$P(j) = \frac{n_j}{N} = (1-\theta)^j \theta \quad (27)$$

For $j = 0, 1, \ldots, T_G$, Eq. (27) is the PMF of the truncated geometric distribution. Therefore, for a regular graph with degree $k \geq 0$, the steady state distribution of tokens over the nodes will follow a truncated geometric distribution. We confirmed this behavior through numerical simulations of two regular graphs, with degrees 0 and 20 (Fig. S1).

When $T_G \gg N$, $\mu = \ln(1 + N/T_G) \approx N/T_G$. Under this condition, the probability distribution in Eq. (26) is equivalent to the Gibbs distribution, with the effective temperature being $T_G/N$, the density of tokens on the graph.

When the degree distribution is not uniform, like in BA and ER graphs, the drift of tokens for a node will vary with its degree. That will change the token distribution at steady state. In the previous section, we have shown that the necessary condition for a node with degree $k_i$ to hold tokens is $k_i \leq \left( \frac{1}{p} - 2 \right)$. When $p < p_c$, nodes with degree $k_i < \left( \frac{1}{p} - 2 \right)$ will have a net positive drift of tokens, and those nodes only will accumulate tokens. Therefore, the minimum-degree nodes will accumulate most of the tokens.

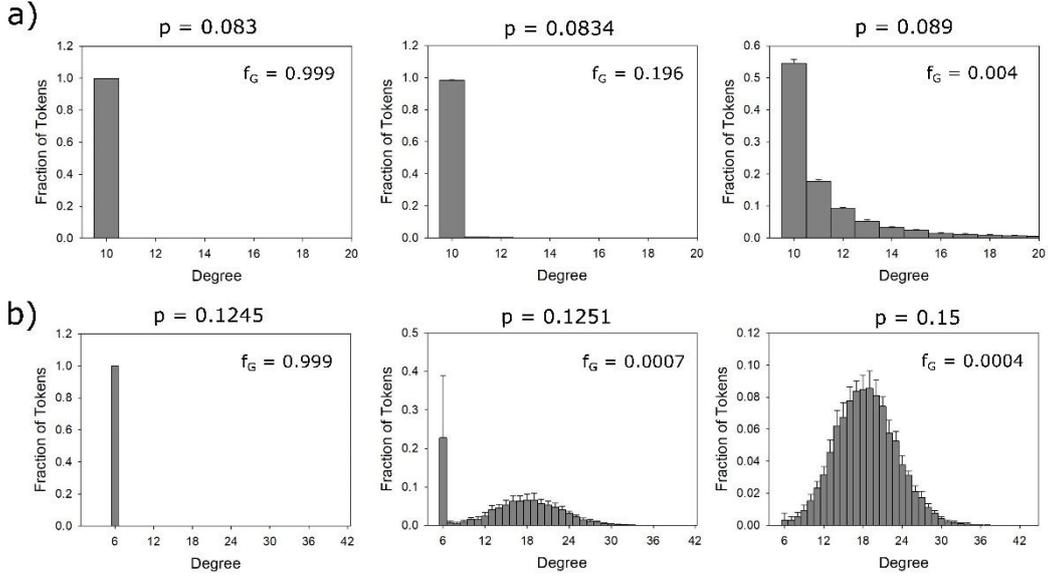

Figure 5. Steady state token distribution over nodes – effect of node degree. a) BA-type graph with $\langle k \rangle = 20$, $k_{min} = 10$, and $p_c = 1/12$. The degree distribution of this graph follows a power law distribution with a long tail. However, the data of higher-degree nodes are not shown here to make the diagram more legible. Those higher degree nodes are empty. b) ER-type graph with $\langle k \rangle = 20$, $k_{min} = 6$, and $p_c = 1/8$. Steady state distributions of tokens are shown for different values of $p$.

We observed the same in our simulations. Fig. 5a shows the token distribution for the BA graph with $k_{min} = 10$. In this case $p_c = 1/12$. The first panel shows the token distribution where $p < p_c$. In this case, all the tokens have accumulated on the nodes of minimum degree. Similarly, in the ER graph with $k_{min} = 6$, at $p < p_c$ all tokens on the graph are accumulated on the minimum degree nodes (Fig. 5b, first panel).

The token distribution on nodes changes when $p > p_c$. In this regime, loss is dominant for all nodes. At steady state, for a $p$ sufficiently above $p_c$, $f_G$ is very small and the nodes are sparsely populated with tokens. As most of the nodes are empty most of the time, almost all attempts to drain a token from a selected node fail, irrespective of its degree. The gain of a token is independent of the node degree in itself. At the steady state, the degree of a node will have a very mild effect on the number of tokens. Therefore, the token distribution will be similar to the degree distribution.

For example, in the BA graph, the degree distribution follows a power law. Nodes of degree $k_{min}$ appear with the highest frequency, and the frequency of nodes decreases with an increase in degree. Therefore, when $p > p_c$, the token distribution should exhibit a similar trend, with the accumulation of tokens decreasing with node degree. We observed this behavior in our simulation (Fig. 5a, $p = 0.089$). Similarly, as the degree distribution of an ER graph follows a Poisson distribution, the token distribution should look like a Poisson distribution. We observed the same in our simulation (Fig. 5b, $p = 0.15$).

The information on the token distribution helps to understand the behavior in the transition zone. Let $p = p_c + \delta$, where $0 < \delta \ll 1$. In this case, the upper bound for the degree of a node that can hold tokens is $k^* = \left(\frac{1}{p} - 2\right) = \left(\frac{1}{p_c + \delta} - 2\right)$. As $k^* < k_{min}$, all nodes will lose tokens. However, the probability of token loss by a node with degree $k_i$ is proportional to $(1 + k_i)$. Therefore, token loss will be less for nodes with a degree close to $k^*$.

In the case of a BA graph, the majority of nodes have the lowest degree $k_{min}$. In the transition zone, where $p$ is very close to $p_c$, $k_{min}$ is very close to $k^*$. Therefore, the majority of the nodes will experience a slow drain of tokens, and the token count should decrease gradually with an increase in $p$. Notice the token distribution for the BA graph in Fig. 5a, for $p = 0.0834$. In this case, $k^* = 9.99$ that is very close to $k_{min} = 10$. The effective rate of loss of tokens from 10-degree nodes would be lower than that of other nodes, and these nodes are the majority on the graph. Therefore, even though token loss is dominant over the gain, the majority of nodes hold some tokens and $f_G$ stays above zero. This behavior leads to a continuous drop in $f_G$ with $p$ in the transition zone (Fig. 3b).

In the case of a regular graph, all nodes have the same degree $k$. Therefore, in the transition zone where $k$ is close to $k^*$, all nodes have the same slow loss of tokens. That leads to a continuous drop in $f_G$ with $p$ (Fig. 3a).

However, in the case of an ER graph, the majority of the nodes have a degree much higher than $k_{min}$. Only a handful of nodes will have a degree close to $k^*$. So, even with a very small $\delta$, tokens drain out of the overwhelming majority of nodes. That leads to a sharp, discrete drop in $f_G$.

Consider $p = 0.1251$ for the ER graph with $\langle k \rangle = 20$ and $k_{min} = 6$. This is very close to $p_c = 0.125$. For this $p$, $k^* = 5.99$. Therefore, nodes with $k = 6$ should have the slowest rate of token loss. However, in our simulations, these nodes are very rare – one or two out of 5000 nodes. Other nodes close to $k^*$ are also very rare. For example, the average number of nodes with degrees 7 and 8 are 2 and 7.67, respectively. As low-degree nodes are an extreme minority, even if they hold some tokens, the total token on the graph is very low at the steady state. That's why the change in $f_G$ is discrete in the transition zone. This behavior can be observed in the degree distribution plot (Fig. 5b, middle panel). At steady state, on average, 0.07% of total tokens are on the graph, out of which most are on the six-degree nodes. The rest are distributed over other nodes.

What would happen at $p_c$? At $p_c$, all nodes except the minimum-degree nodes have a negative drift. Therefore, they cannot hold tokens, and as a result, they do not make significant contributions to the net token dynamics. However, the drift is zero for the nodes with the minimum degree. For these nodes, token exchange is equivalent to a random walk, and the behavior should be the same as we derived for the graph without edges at $p = 0.5$. Following this logic, we expect $f_G = N_{min}/(N_{min} + 1)$ at $p_c$. Here $N_{min}$ is the number of minimum-degree nodes. Therefore, when the number of minimum-degree nodes is very large, $f_G \approx 1$ at $p_c$, and tokens accumulate on the minimum-degree nodes. We have observed this for both regular and BA graphs.

However, in an ER-type graph, the number of the minimum-degree nodes is very low. That has two consequences. At $p_c$, almost all nodes have a net negative drift. Whenever these nodes gain a token, they quickly lose it. As they go empty, most attempts to remove tokens from nodes fail. Effectively, most of the steps in the simulation do not yield any net change in the number of tokens on the graph. That slows down token dynamics immensely, similar to the phenomenon of critical slowing down observed in second-order phase transitions and bifurcation.

Extreme low abundance of minimum-degree nodes also affects $f_G$ at $p_c$ in the ER graphs. As $N_{min}$ is not very high, we can not do the approximation that $f_G = N_{min}/(N_{min} + 1) \approx 1$. Therefore, any ER graph with a reasonable number of nodes will have $f_G = N_{min}/(N_{min} + 1) < 1$ at $p_c$.

Due to the extremely slow dynamics, we could not investigate the steady state behavior of ER graphs with 5000 nodes exactly at $p_c$. However, we investigated the behavior for a smaller ER graph with 100 nodes, $\langle k \rangle = 8$ and $k_{min} = 2$. For this graph $p_c = 0.25$. We modified the simulation algorithm to generate ER graphs with a specific number of minimum degree nodes, $N_{min}$. As shown in Fig. S2, for this graph $f_G$ behaves in the expected way. For, $N_{min} = 1$ the $f_G$ at $p_c$ is $0.57 \pm 0.19$. When $N_{min}$ is increased to 5, $f_G$ at $p_c$ moved closer to 1 ($f_G = 0.84 \pm 0.05$).

In summary, we observed two crucial steady state behaviors. When the degree distribution is not uniform, in the token-saturated state, $f_g \approx 1$, all tokens accumulate or condense on the minimum-degree nodes of the graph. The number of the minimum degree nodes also dictates the transition behavior when $p > p_c$. When the minimum-degree nodes are very high in number, $f_G$ falls sharply to zero, but in a continuous fashion, as the minimum-degree nodes act like a buffer holding some tokens. However, when the minimum-degree nodes are present in extremely low numbers (as in an ER graph), $f_G \approx 1$ only for $p < p_c$. At $p_c$, $f_G$ drops to a value lower than one, and further increase in $p$ leads to a sharp discrete drop in $f_G$ towards zero.

### D. Dependence on system size and number of tokens

For all graphs, when $N$ and $T$ are finite and $T \gg N \gg 1$, $f_G \approx 1$ for $p < p_c$. However, the transition from $f_G \approx 1$ to 0, depends upon the value of $N$ and $T$. So, we investigated the finite-size effect of $N$ and $T$ on $f_G$ when $p > p_c$.

For an edgeless graph, using Eq. (19), we get the following scaling relation, when $p > p_c$ and $N \gg 1$,

$$f_G = \left(\frac{N}{T}\right) F_G(p - p_c) \quad (28)$$

As, for this graph
$p_c = 0.5$, $F_G(p - p_c) = 0.25/(p - p_c) - 0.5$.

Note that when $p$ is very close to 0.5, $(1-p)/(2p-1)$ in Eq. (19) explodes and tends to an asymptote. In such a case, Eq. (19) may give $f_G > 1$, which is not possible. However, if $T/N$ is sufficiently large to counter $(1-p)/(2p-1)$ then $f_G < 1$. Therefore, the scaling relation in Eq. (28) is appropriate only when $T/N$ is sufficiently larger than $(1-p)/(2p-1)$.

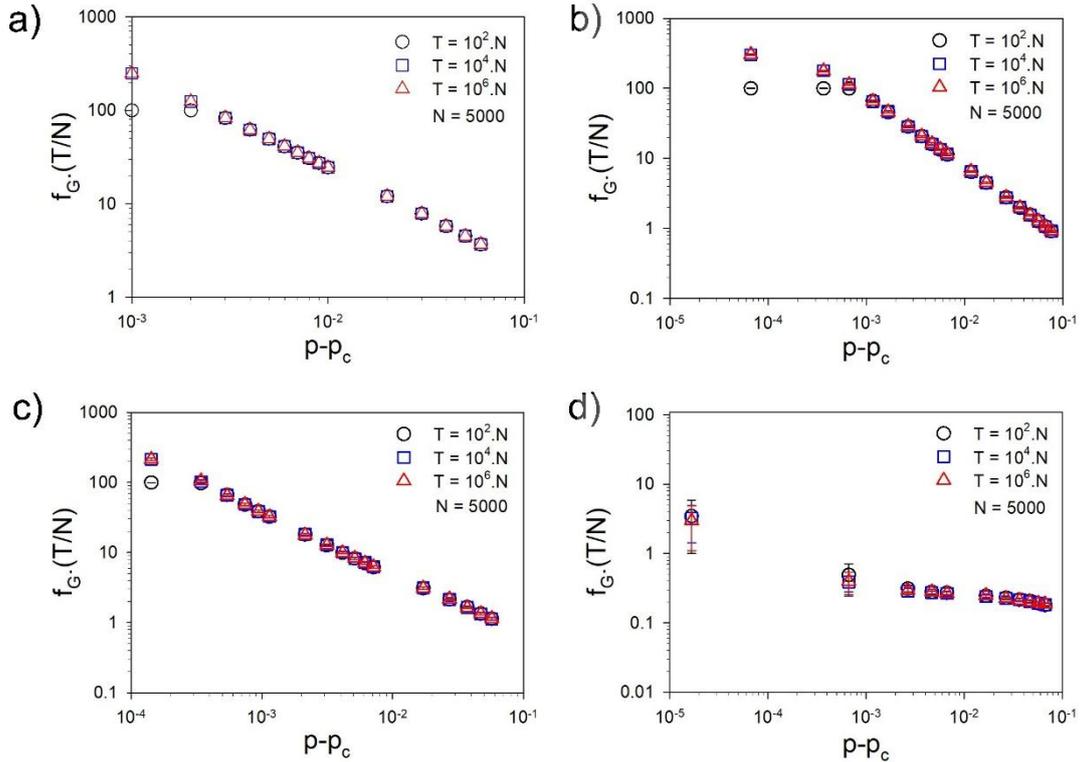

Figure 6. Scaled behavior of $f_G$ for different values of $T$ in the transition zone. Shown in log-log plots. a) edgeless graph, b) regular graph ($k = 10$), c) BA-type graph ($\langle k \rangle = 10$), and d) ER-type graph ($\langle k \rangle = 10$, $k_{min} = 1$). Data only for $p$ values very close to $p_c$ are shown to highlight the rapid fall in $f_G$. The data in (a) was generated by numerically evaluating Eq. (11). MC simulations were used for graphs in (b) to (d) and averages of multiple independent simulations are shown. In most cases the error bars are smaller than the symbols.

We calculated $f_G$ for an edgeless graph with $N = 5000$ using Eq. (11) for three different, widely separated values of $T$. As shown in Fig. 6a, the data for these three $T$ values collapsed on each other in $f_G(T/N)$ vs $(p - p_c)$ plot.

As expected for the two lowest values of $p$, the data for $T = 10^2 N$ do not collapse with other data sets, as $T/N$ is not sufficiently larger than $(1-p)/(2p-1)$.

The steady state behavior for a connected graph is different from the edgeless graph only in two ways – the rate of loss of tokens for a node depends upon its degree, and the critical threshold depends on the minimum degree of the graph. Otherwise, all three connected random graphs show a sharp, switch-like transition in the fraction of tokens on the graph. Therefore, we expect that these graphs will also show the scaling relation in Eq. (28). However, we can not predefine the function $F_G(p - p_c)$. We have discussed earlier that the $p_c$, and the behaviour of $f_G$ in the transition zone depends upon the graph topology, particularly on $k_{min}$ and $N_{min}$. Therefore, $F_G(p - p_c)$ would depend upon the graph topology.

Fig. 6b, shows the $f_G(T/N)$ vs $(p - p_c)$ plot for a regular graph with a fixed $N$ but widely separated values of $T$. From the observed data collapse in this figure, it is evident that the scaling relation is applicable for a regular graph. Similar simulations were performed for BA and ER graphs. It is evident from Fig. 6c and d, that these two graphs also follow the same scaling relation.

In summary, our scaling analysis shows that, for $p > p_c$, all dependence of $f_G$ on the extensive variables $T$ and $N$ enters only through their ratio $T/N$. This ratio is an intensive parameter, and it controls the sharpness of the switching transition.

This observation raises another interesting question. What would be the steady state behavior of our model for a very large but finite size system ($N \gg 1$) with $T \to \infty$? That is equivalent to a large system in contact with an environment that has an enormous supply of tokens, as is often the case in many real-life systems.

It is obvious that $f_G$ will have the same switch-like transition beyond $p_c$. However, the transition behaviour will be affected. For a finite but large $N$, when $T \to \infty$, $N/T \to 0$. For $p > p_c$, following Eq. (28), $f_G \to 0$. Therefore, in the limit of $T \to \infty$, with finite N, the switching transition becomes effectively discontinuous with $f_G$ collapsing sharply to zero for $p > p_c$.

## IV. DISCUSSION

In this work, we investigated a stochastic token-exchange model on a graph. In this model, tokens are exchanged between the nodes of the graph and a reserve. However, the token exchange is not symmetric. At every step, a randomly selected node can gain one token with probability $(1-p)$; whereas with probability $p$, the selected node and its neighbors can lose one token each. So, at one step, the graph can gain one token but lose more than one. This collective loss of tokens in a neighborhood breaks detailed balance. Furthermore, the total number of tokens (in the reserve plus on the graph) is constant. This conservation rule imposes a global constraint on the system, which otherwise evolves through the local exchange of tokens between individual nodes and the reserve.

We used the fraction of tokens on the graph at steady state ($f_G$) as the order parameter and $p$ as the control parameter. $p$ regulates the drift of tokens between the graph and the reserve. We investigated the steady state behavior of this system for different graph topologies – a graph without edges, regular graph, Barabási-Albert (BA), and Erdős–Rényi graphs (ER).

We have shown that for a large graph, the order parameter $f_G$ undergoes a sharp switch-like change as the control parameter $p$ crosses a critical threshold $p_c$. When $p < p_c$ almost all the tokens accumulate on the graph. Beyond $p_c$, the graph loses tokens drastically and $f_G$ drops sharply to zero. So, our model has a state transition with two states− a token saturated state ($f_G \approx 1$) to an empty state ($f_G \approx 0$).

This switch-like transition is not the result of a connection between nodes, as a graph without edges also shows the transition in the limit of $N \gg 1$. This transition is a result of the large system size and the conservation of total tokens. When $p < p_c$, nodes with degree $k_i < (1/p - 2)$ accumulate tokens. As the token number is constant, the reserve will eventually be completely drained. Similarly, with a small increase in $p$ above $p_c$, all the nodes simultaneously lose tokens, triggering a sharp drop in $f_G$.

However, the critical threshold of this transition is decided by the topology of the graph, and we show that $p_c = 1/(k_{min} + 2)$. Here $k_{min}$ is the minimum degree of the graph.

The topology has another effect on the steady state behavior. The probability of token loss depends upon the node degree. Therefore, low-degree nodes will always have a higher likelihood of accumulating tokens. However, the frequency of such low-degree nodes depends on the graph topology. For example, in a BA graph, nodes with $k_{min}$ degree are the most abundant ones. However, in the ER graph, they are the least abundant. In some cases, their counts drop to one or two. The abundance of low-degree nodes, particularly the node with $k_{min}$ degree, decides how many tokens can be on the graph at the steady state. That in turn decides the behavior of $f_G$ in the transition zone, where $p$ is slightly bigger than $p_c$ and $f_G$ drops sharply.

In a BA graph, the overwhelming majority of nodes have the lowest degree, and they tend to hold tokens even in the transition zone. Therefore, in the transition zone, $f_G$ decreases smoothly with a change in $p$. Regular graphs with degree $k \geq 0$ have the same behavior. However, as the lowest-degree nodes are an extreme minority in an ER graph, a slight increase in p drains the graph empty. That leads to the drop in $f_G$ through discrete jumps.

In the token-saturated state ($f_G \approx 1$), for a graph with non-uniform degree distribution, tokens accumulate on the lowest degree node. That is similar to condensation observed in the zero-range process (ZRP) [20]. In ZRP, a finite and fixed number of particles hop stochastically among a finite number of sites, and the hopping rate from a site depends only on its occupation number. In the inhomogeneous ZRP, sites differ in their intrinsic hopping parameter. Therefore, the hopping rate from a site depends upon both the identity of the site and the number of particles on it. When the particle density exceeds a critical threshold, particles condense on the site with the lowest hopping rate [21]. That site holds a macroscopic number of particles, while other sites hold a negligible number of particles. Similarly, in our model, the minimum-degree nodes have the least chance of losing tokens, and when $p < p_c$ the tokens accumulate on those nodes.

This behavior is extreme in the case of an ER graph, as the number of minimum-degree nodes is extremely low. In an ER graph, when $p < p_c$, all the tokens accumulate on those few nodes. However, when $p > p_c$ only a few tokens are retained in the graph, they get distributed sparsely over nodes with different degrees. Such an extreme change in token distribution with a change in $p$ is akin to the condensation-transition of ZRP.

In brief, our model is a minimal model of drift-driven nonequilibrium state transition of a system with open, interconnected subsystems that exchange resources with a common reserve. Each subsystem experiences a stochastic drift that determines the gain or loss of resources, and this drift is regulated by a system-wide control parameter ($p$ in our model). This control parameter need not be an externally regulated one. It could be a property of the system itself.

As the control parameter is varied, the system undergoes a sharp switch-like state transition between the high- and low-resource states. The switch-like transition arises not from explicit subsystem interactions but from the interplay of drift and conservation. There is no thermodynamic drive either—no defined energy function or energy minimization principle guiding the dynamics. The dynamics is purely stochastic, governed by asymmetric transition probabilities that regulate the distribution of a conserved quantity.

However, the critical threshold and the behaviour in the transition zone are determined by the internal architecture of the system, particularly by the frequency of the least connected subsystems and their connectivity. The internal connectivity also affects the distribution of resources among subsystems and can lead to the extreme accumulation of resources in a few subsystems.

Taken together, our model demonstrates that a simple drift–conservation mechanism, with a heterogeneous subsystem interconnection, is sufficient to generate diverse behaviors—from condensation-like accumulation to sharp, switch-like state transitions. We expect that similar mechanisms may operate in different physical, biological, and social systems where interconnected subsystems exchange a conserved resource with a common environment.

# Supplementary Information

**Switching Transition in a Resource Exchange Model on Graphs**
Shreeman Auromahima, Sitangshu Bikas Santra, Biplab Bose

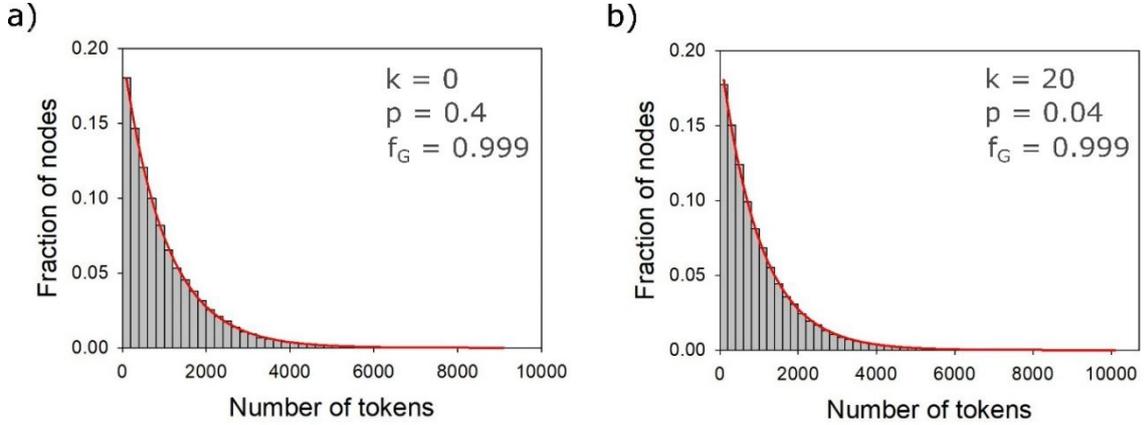

Figure S1: Steady state distribution of tokens on regular graphs. a) for a graph without any edges ($k = 0$) and b) for a graph with degree $k = 20$. In both cases, the data on the number of tokens on each node was binned into equal-sized bins and then displayed as a frequency histogram. The pattern of distribution in both cases follows a geometric distribution. It can be shown that the binned data from a geometric distribution $P(j) = (1-p)^j p$ fits the exponential function $A\exp(\ln(1-p)x_i)$ where $x_i$ is the bin-centre of the $i$-th bin. The red lines represent the fitted exponential functions. In both cases, $N = 5000$ and $T = 1000N$.

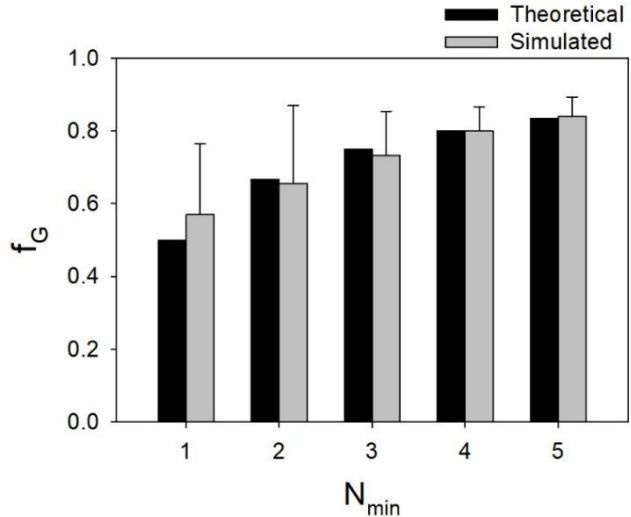

Figure S2: Behaviour of ER graphs at $p_c$. The theoretical and simulated values of $f_G$ at $p_c$ for different $N_{min}$. Here $N_{min}$ is the number of nodes with the minimum degree $k_{min}$. As discussed in the main text, at $p_c$ the theoretical value of $f_G = N_{min}/(N_{min}+1)$. For simulations, ER graphs with $N = 100$, $T = 1000N$, $\langle k \rangle = 8$, and $k_{min} = 2$ were used. The algorithm was modified such that graphs with a specified $N_{min}$ are generated and used for simulation. For all these graphs $p_c = 0.25$.